\documentclass[11pt, a4paper]{article}
\usepackage[utf8]{inputenc}
\usepackage[T1]{fontenc}
\usepackage{graphicx}
\usepackage{listings}
\usepackage{color}
\usepackage{amssymb, amsmath, geometry}
\usepackage{bbm}
\usepackage[numbers]{natbib}
\usepackage{hyperref}
\usepackage{setspace} 
\usepackage{booktabs}
\usepackage{caption, subcaption}
\usepackage{authblk}

\newcommand{\tp}[1][3]{^{^{\mkern-#1mu\mathsf{T}}}}

\newcommand{\rvline}{\hspace*{-\arraycolsep}\vline\hspace*{-\arraycolsep}}
\newcommand{\E}{{\rm E}}

\begin{document}

\title{Competing risks models with two time scales}
\author{}
\date{}
\author{Angela Carollo\authorcr\small
        \texttt{carollo@demogr.mpg.de}\authorcr
        Max Planck Institute for Demographic Research\authorcr
        Rostock, Germany \authorcr
        Leiden University Medical Center, Leiden, The Netherlands
        \and
        Hein Putter\authorcr\small
        \texttt{h.putter@lumc.nl}\authorcr
        Leiden University Medical Center, Leiden, The Netherlands
        \and
        Paul H.C.~Eilers\authorcr\small
        \texttt{p.eilers@erasmusmc.nl}\authorcr
        Erasmus University Medical Center, Rotterdam, The Netherlands
        \and
        Jutta Gampe\authorcr\small
        \texttt{gampe@demogr.mpg.de}\authorcr
         Max Planck Institute for Demographic Research\authorcr
        Rostock, Germany
        }

\maketitle

\begin{abstract}
% Currently 191 words.
\noindent  
Competing risks models can involve more than one time scale. A relevant example is the study of mortality after a cancer diagnosis, where time since diagnosis but also age may jointly determine the hazards of death due to different causes. Multiple time scales have rarely been explored in the context of competing events. Here, we propose a model in which the cause-specific hazards vary smoothly over two times scales. It is estimated by two-dimensional $P$-splines, exploiting the equivalence between hazard smoothing and Poisson regression. The data are arranged on a grid so that we can make use of generalized linear array models for efficient computations. The R-package \texttt{TwoTimeScales} implements the model.

As a motivating example we analyse mortality after diagnosis of breast cancer and we distinguish between death due to breast cancer and all other causes of death. The time scales are age and time since diagnosis. We use data from the Surveillance, Epidemiology and End Results (SEER) program. In the SEER data, age at diagnosis is provided with a last open-ended category, leading to coarsely grouped data. We use the two-dimensional penalised composite link model to ungroup the data before applying the competing risks model with two time scales.

\end{abstract}
\vskip 3ex

\newpage

\section{Introduction}
\label{sect:intro}
Competing risks describe the situation where individuals are at risk of experiencing one of several types of events. \cite{Putter:2007} The prototype of a competing risks model is the study of cause-specific mortality. Such models are used in clinical studies of cancer to analyse mortality from cancer as well as mortality due to other causes as competing events.

The building blocks in a competing risks analysis are the cause-specific hazards. They are defined as the instantaneous risk of experiencing an event of a specific type at time $t$, given no event (of any type) has happened before $t$. From them the overall survival function, i.e., the probability of no event up to time $t$, and the cumulative incidence functions, the probability of an event of given type before $t$, can be derived. 

Time is a key quantity in any survival analysis, and it can be recorded on several time scales. For example, after a cancer diagnosis, the risk of death may be studied over time since diagnosis, over age, which is time since birth, or over time since treatment. 
All time scales progress at the same speed and differ only in their origin. However, each time scale is a proxy for a specific mechanism linked to the event of interest. \cite{Berzuini:1994} In the study of mortality after a cancer diagnosis, time since diagnosis measures the cumulative adverse effect of the cancer. Additionally, as individuals age, they become more frail and their capacity of resisting comorbidity deteriorates.

Usually, cause-specific hazards are defined for the same single time scale, and little research has been done on how to handle multiple time scales in competing risks models. Existing studies focus on the choice of the time scale to use when modeling competing causes of deaths after a cancer diagnosis. Cancer mortality is clearly a function of time since diagnosis, while other causes of death are more naturally modeled over age. Whenever the cause-specific hazards of death for other causes are solely modeled over the time since diagnosis, bias may be introduced in the estimates of the cumulative incidences for cancer mortality. \cite{Skourlis:2022} Lee et al. \cite{Lee:2016} propose a model where two competing causes are modeled on two different time scales and then combined under one of the two scales to estimate the cumulative incidence functions. 

Carollo et al. \cite{Carollo:2023} introduced a model for a single event in which the hazard varies smoothly over two time scales simultaneously and is estimated by tensor products of $P$-splines. \cite{Currie:2004} The model allows great flexibility for the hazard shape, while being simple to estimate. Here, we develop this model further for a competing risks setting. Each cause-specific hazard varies over the two time scales, and estimation again is achieved by two-dimensional $P$-splines smoothing. Therefrom, we calculate the cumulative incidence functions for each cause.

This study is motivated by the analysis of mortality after a breast cancer diagnosis. Breast cancer is the most common cancer diagnosed in women worldwide, \cite{Arnold:2022} and the risk of being diagnosed with it increases over age, but varies by race and ethnic group, \cite{Stapleton:2018} with peaks of diagnosis after age 60 for white women and in the late 40s for non-white women. Survival probabilities after a breast cancer diagnosis are generally very high, with more than 80\% of women with breast cancer being alive 10 years after diagnosis, \cite{SEERexplorer:BCsurv} leading to many women living with a history of breast cancer for many years. Several studies have shown that breast cancer survivors are at higher risk of dying from cardiovascular disease (CVD), \cite{Colzani:2011, Bradshaw:2016, Mehta:2018} and there is evidence that this increased risk appears among survivors of breast cancer several years after diagnosis. \cite{Bradshaw:2016} Increased CVD mortality rates among breast cancer survivors appear to be likely linked to adverse effects of therapy and to common risk factors. \cite{Mehta:2018}

Mortality rates of women with breast cancer depend on age, race and cancer subtypes, \cite{Brandt:2015, Jayasekara:2019} however, the relationship between time since diagnosis, age, age at diagnosis and mortality is not yet clear. Most studies of women with breast cancer find that breast cancer mortality increases after age 70 so that older women are at higher risk of dying because of the cancer. \cite{SanMiguel:2020, Freedman:2018, Brandt:2015} However, older women with breast cancer often die of causes other than cancer, with cardiovascular diseases being the most common cause of non-cancer death.\cite{Mehta:2018} Other studies have found that younger ages at diagnosis of breast cancer are associated with higher overall mortality \cite{Jayasekara:2019, Partridge:2016} and that women diagnosed before age 35 had higher risk of death because of cancer than older women. \cite{Narod:2015} Although most studies acknowledge the presence of several time scales, they do not account for them explicitly. For example, age at diagnosis is often grouped in broad and arbitrarily-cut categories, and interaction with time since diagnosis is often neglected.
Shedding light on the way mortality rates of breast cancer vary with these two time scales, while properly accounting for competing causes, is both of empirical and of methodological relevance.

We analyse mortality data of women with breast cancer from the Surveillance, Epidemiology and End Results (SEER) registers \cite{SEER:program} by time since cancer diagnosis and age. We focus on postmenopausal women (age at diagnosis 50 and older) and distinguish between deaths due to breast cancer and all other causes of death. In the SEER data, age at diagnosis is recorded in single years of age up to age 89 and one last open-ended interval for all individuals older than 89 (age at diagnosis 90+). Such grouping is common in data on disease incidence and/or mortality, usually to prevent identification of the patients or to provide a condensed depiction of the data. In this paper we employ the two-dimensional penalised composite link model (PCLM) \cite{Rizzi:2018} to ungroup the final age interval.

The remainder of this article is structured as follow: In Section~\ref{sect:method} we first set the notation for the cause-specific hazards with two time scales and review the estimation procedure. Then we introduce calculation of cumulative cause-specific hazards, overall survival probabilities and finally cumulative incidence functions with two time scales, together with their standard errors. We also discuss the two-dimensional PCLM to ungroup the data in the final age interval on two time scales. Section~\ref{sect:application} is dedicated to the motivating example. A discussion concludes the paper.

\section{A competing risks model with two time scales}
\label{sect:method}

Consider individuals that are diagnosed with cancer. Once a patient is diagnosed with the disease, he or she is at risk of dying from cancer (event of interest) or because of causes other than cancer (competing event). 
The risk of dying from cancer actually sets in at disease onset, however, since this onset is rarely known, time since diagnosis mainly is used as duration of the disease and we will also follow this practice.

For both competing events two time scales are relevant: The first is the age of the individual, which we indicate with $t$. The second time scale is time since diagnosis of the cancer, which we denote with $s$. Because age measures the time since birth, we have that $t > s$ for each individual. Both $t$ and $s$ move at the same speed, that is, an increment of one unit (for example one day or one month) on the $t$-scale corresponds to the same increment on the $s$-scale. 

We consider two competing events: Death from beast cancer ($\ell = 1$) and death from other causes ($\ell = 2$). The cause-specific hazards (CSHs) for event type $\ell \in\{1, 2\}$ over the two time scales $t$ and $s$ are defined as
\begin{equation}
\label{eq:csh}
\lambda_\ell(t,s) = \lim\limits_{\nu \downarrow 0}\frac{ \text{P} \left ( \; \text{event of type }\ell \in \{ t+\nu,s+\nu\} \,|\,
\text{no event of any type before } (t,s) \; \right ) } {\nu}.
\end{equation}
Here $t > s$, so the two-dimensional hazards $\lambda_{\ell}(t,s)$ are only defined in the lower half-open triangle of $\mathbb R_+^2$. They give the instantaneous risk of dying because of cause $\ell$ for an individual who is alive at age $t$ and $s$ years after receiving the diagnosis of cancer.

The two time scales differ in their origin, which in the case of cancer patients is the age $t_0$ when the individual is diagnosed with cancer. This age is fixed for each individual but differs among them. In a Lexis diagram with axes $t$ (age) and $s$ (time since diagnosis) individuals move along diagonal lines from $(t_0, 0)$ to $(t_0+v, v)$ until they leave the risk set (due to event of any kind or censoring). The individual trajectories start at individually different points $(t_0, 0)$.

The same information can also be portrayed in the $(u,s)$-plane, where $u$ denotes the age at diagnosis and $s$, as before, is the time since diagnosis. Here individual trajectories run vertically from $(u, s=0)$ to $(u, s=v)$. We can view the cause-specific hazards equivalently as two-dimensional functions of $u$ and $s$, $\breve\lambda _\ell (u,s)$, where 
\begin{equation}\label{eq:2Dspec}
\breve\lambda _\ell (u=t-s,s) \equiv \lambda_\ell(t,s), 
\end{equation}
(see Carollo et al.\cite{Carollo:2023}). The $\breve\lambda _\ell (u,s)$ are defined over the full positive quadrant $\mathbb R^2_+$.

From the CSHs $\lambda_\ell(t,s)$ or $\breve \lambda _\ell(u,s)$, respectively, we obtain the cumulated cause-specific hazards, the overall survival probability and the cumulative incidence functions. In the next section we describe estimation of the CSHs. Thereafter, we provide expressions for the quantities that are obtained from the them.

\subsection{Cause-specific hazards: Model and estimation}
\label{subsect:estimation}

We will model (and estimate) the cause-specific hazards $\breve\lambda _\ell (u,s)$ over the $(u,s)$-plane. Back-transformation to the $(t,s)$-coordinates is straightforward using Equation (\ref{eq:2Dspec}). The only assumption that we are going to make about the cause-specific hazards is that they  vary smoothly over $u$ and $s$ (and hence over $t$ and $s$). This will be achieved by two-dimensional $P$-spline smoothing (tensor products of $B$-splines with difference penalties on the rows and columns of coefficients, see Eilers \& Marx \cite{EilersMarx:2021}). The approach was introduced by Carollo et al. \cite{Carollo:2023} for a single event type and we will extend it here to the competing risks setting.

The $(u,s)$-plane is divided into $n_u\times n_s$ small bins of size $h_u$ and $h_s$, respectively (rectangles if $h_u \neq h_s$ and squares otherwise), covering the range of observed values of $u$ and $s$. The bin widths $h_u$ and $h_s$ will be chosen relatively narrow and hence the number of bins $n_u$ and $n_s$ can be relatively large.
Within each bin the number of events of type $\ell$, denoted by $y^{\ell}_{jk}$, and the total time at risk $r_{jk}$ are determined ($j=1,\ldots, n_u$; $k=1,\ldots, n_s$), summed over all individuals~$i, i=1,\ldots, n$. (In the following the superscript $\ell$, like in $y^{\ell}_{jk}$, indicates the cause and not a power.)

Each individual $i$ contributes to $y^{\ell}_{jk}$ only in bins with a single index $j$, namely, where its age at diagnosis~$u_i$ is located. If the individual experiences an event of type $\ell$, then it contributes 1 to the event count $y^{\ell}_{jk}$ in the bin $(j,k)$, in which its exit time (since diagnosis) $s_i$ falls. In all other bins, and in all bins of the other cause it contributes 0. For an individual that is right-censored (leaves the risk set without having experienced any of the events) all contributions to the $y^{\ell}_{jk}$ equal zero. 

The contribution of an individual's at-risk time per bin is, again, only in those cells where the first index~$j$ is determined by the individual age-at-diagnosis $u_i$, and the contributions are as long as the individual is under observation in the respective bin. Exit from the risk set is caused by experiencing an event of any type or by right-censoring. Late entry can be accommodated in a straightforward way in that an individual contributes only to those bins, where it is in the risk set (and not automatically from $s=0$ in $r_{j1}$). In our application all individuals are followed-up from diagnosis onward, so there is no late entry. The $r_{jk}$ are the sum over all individual exposure times in the respective bin.

The $y^{\ell}_{jk}$ can be thought of as realizations of Poisson variates 
\cite{Holford:1980, Laird:1981} with means 
\begin{equation}
\label{eq:Poisson}
\breve \mu ^{\ell}_{jk} = r_{jk} \, \breve \lambda ^{\ell}_{jk} = 
r_{jk} \, \exp \left ( \breve \eta ^{\ell}_{jk} \right ),
\end{equation}
where the $\breve \lambda ^{\ell}_{jk} $ represent the cause-specific hazards $\breve\lambda_\ell(u,s)$ evaluated at the center of bin $(j,k)$. The  $\breve \eta ^{\ell}_{jk}$ are the corresponding values of the log-hazard, $\breve \eta ^{\ell}_{jk} = \ln \breve \lambda ^{\ell}_{jk} $.

Because of the same-sized bins, event counts and at-risk times are on a regular grid and are naturally arranged as $n_u\times n_s$ matrices 
$\mathbf{Y}_\ell=[ y^{\ell} _{jk} ]$ % \text{\qquad and \qquad}
and 
$  \mathbf{R} =[r_{jk}] $.
Correspondingly, we denote $\mathbf{M}_\ell=[\breve \mu ^{\ell} _{jk}]$ and  $\mathbf{E}_\ell=[\breve \eta ^{\ell} _{jk}] $, so that Equation (\ref{eq:Poisson}) can be written more concisely in matrix form as 
\begin{equation}\label{eq:Poisson2}
\mathbf{Y}_\ell \sim \text{Poisson} (\mathbf{M}_\ell) \text{\qquad with \qquad} 
\mathbf{M}_\ell = \mathbf{R} \odot \exp \left ( \mathbf{E}_\ell \right ) .
\end{equation}
Here $\odot$ denotes element-wise multiplication.

To obtain smooth cause-specific hazard surfaces we model the log-hazards $\breve\eta_\ell(u,s) = \ln \breve\lambda_\ell(u,s)$ via tensor products of $B$-splines. Modeling the log-hazard will automatically produce positive estimates for the cause-specific hazards. 

The tensor products are formed from two marginal $B$-splines bases along the $u$- and $s$-axis, with $c_u$ and $c_s$ elements, respectively. We choose cubic $B$-splines along both axes. The two basis matrices are denoted by $\mathbf{B}_u$, which is $n_u \times c_u$, and by $\mathbf{B}_s$, which is $n_s \times c_s$. The $c_u c_s$ regression coefficients are denoted by $\alpha^{\ell}_{lm}$ ($l = 1,\dots,c_u$; $m=1,\dots,c_s$), and the log-hazards, evaluated at the bin midpoints, are
\begin{equation}\label{eq:eta}
\breve \eta_{jk}^{\ell} = \sum_{l=1}^{c_u}\sum_{m=1}^{c_s} b_{jl}^u b_{km}^s \alpha_{lm}^{\ell}.
\end{equation}
If we arrange the coefficients in the $c_u \times c_s$ matrix $\mathbf{A}_\ell=[\alpha^{\ell}_{lm}]$, then the linear predictor in (\ref{eq:eta}) can be expressed in matrix form as
$ \mathbf{E}_\ell = \mathbf{B}_u\mathbf{A}_\ell \mathbf{B}_s\tp$.

The numbers of basis functions $c_u$ and $c_s$ will be large enough to allow sufficient flexibility, but difference penalties on the coefficients, both along the rows and the columns of $\mathbf{A}_\ell $, will prevent over-fitting. The penalty on the coefficients is constructed from two matrices $\mathbf{D}_u$ and $\mathbf{D}_s$ that form differences of order $d$ (usually $d=1$ or $d=2$) of neighboring elements in the columns of a matrix and it is controlled by two smoothing parameters $\varrho_u$ and $\varrho_s$ to allow anisotropic smoothing:
\begin{equation}\label{eq:penalty}
\mbox{pen} (\varrho_u, \varrho_s) = \varrho_u \, || \mathbf{D}_u \mathbf{A}_\ell||^2_{\text{\tiny F}} + \varrho_s \, ||\mathbf{A}_\ell \mathbf{D}_s \tp||_{\text{\tiny F}}^2 
\end{equation}
(The Frobenius norm $||.||^2_{\text{\tiny F}}$ is the sum of all squared elements of a matrix.) 

The objective function to be minimized is the sum of the Poisson deviance resulting from~(\ref{eq:Poisson2}) and the above penalty~(\ref{eq:penalty})
\begin{equation}
\label{eq:pdeviance}
\begin{split}
 \text{dev}(\mathbf{M}_\ell; \mathbf{Y}_\ell) + \mbox{pen} (\varrho_u, \varrho_s) =  2 \sum_{j=1}^{n_u}\sum_{k=1}^{n_s}\left(y^{\ell}_{jk} \ln(y^{\ell}_{jk} / \breve \mu ^{\ell}_{jk}) - (y^{\ell}_{jk} - \breve \mu ^{\ell}_{jk})\right) + \mbox{pen} (\varrho_u, \varrho_s) ,
 \end{split}
\end{equation}
which leads to normal equations that can be solved, for given $\varrho_u $ and $\varrho_s$, in a penalized Poisson Iterative Weighted Least Squares (IWLS) scheme (in compact notation; $\otimes$ denotes the Kronecker product):
\begin{equation}
\label{eq:piwls}
\left[(\mathbf{B}_s \otimes \mathbf{B}_u)\tp \tilde{\mathbf{W}}_\ell (\mathbf{B}_s \otimes \mathbf{B}_u) + \mathbf{P}\right] {\boldsymbol{\alpha}}_\ell  = (\mathbf{B}_s \otimes \mathbf{B}_u)\tp \tilde{\mathbf{W}}_\ell \tilde{\mathbf{z}}_\ell.
\end{equation}
In the penalty matrix $\mathbf{P} = \rho_u (\mathbf{I}_s \otimes \mathbf{D}_u\tp \mathbf{D}_u) + \rho_s (\mathbf{D}_s\tp \mathbf{D}_s \otimes \mathbf{I}_u)$ the  $\mathbf{I}_u$ and $\mathbf{I}_s$ are identity matrices of appropriate dimension. 
The $c_u c_s \times 1$ vector $\boldsymbol{\alpha} _\ell$ holds the elements of $\boldsymbol{A} _\ell$, and the 
$n_un_s \times n_un_s$ matrix $\mathbf{W}_\ell$ is a diagonal matrix of weights (for the Poisson case $\mathbf{W}_\ell=\text{diag}(\boldsymbol \mu _\ell)$). 
The $\mathbf{\tilde z}_\ell$ is the usual working variable $\mathbf{\tilde z}_\ell =\boldsymbol \eta_\ell + (\mathbf y_\ell -\boldsymbol\mu_\ell)/\boldsymbol\mu _\ell$. 
The tilde indicates the current value in the iteration.

The parameters $\varrho_u$ and $\varrho_s$ control the smoothness of the estimated CSHs. Their optimal values are selected by minimizing AIC (Akaike Information Criterion \cite{Akaike:1974}) or BIC (Bayesian Information Criterion \cite{Schwarz:1978}). Both criteria balance fidelity of the estimated model (indicated by $\widehat{\mathbf{M}}_\ell$ and 
$\widehat{\mathbf{W}}_\ell$ below) to the data, as measured by the deviance, and model complexity:
\begin{equation}\label{eq:AICBIC}
\text{AIC}(\varrho_u, \varrho_s) = \text{dev}(\widehat{\mathbf{M}}_\ell; \mathbf{Y}_\ell) + 2 \,\text{ED} \qquad\qquad \text{BIC}(\varrho_u, \varrho_s) = \text{dev}(\widehat{\mathbf{M}}_\ell; \mathbf{Y}_\ell) + \ln{} (n_{\text{bin}}) \,\text{ED} .
\end{equation}
The effective dimension ED of the estimated model is obtained as trace of the hat matrix 
\begin{equation*}
    \text{ED} = \text{trace} \{\mathbf{B} (\mathbf{B}\tp \widehat{\mathbf{W}}_\ell \mathbf{B} + \mathbf{P})^{-1} 
\mathbf{B}\tp \widehat{\mathbf{W}}_\ell \} = 
\text{trace} \{ (\mathbf{B}\tp\widehat{\mathbf{W}}_\ell \mathbf{B} + \mathbf{P})^{-1} 
\mathbf{B}\tp \widehat{\mathbf{W}}_\ell \mathbf{B}\}
\end{equation*}
and $n_{\text{bin}}$ is the number of Poisson variates entering in model (\ref{eq:Poisson2}). 
AIC penalizes model complexity less strongly and has been found to undersmooth, particu\-larly for large sample sizes.\cite{Currie:2004, Camarda:2012} We determine the optimal values for $\varrho_u$ and $\varrho_s$ by numerically minimizing $\text{BIC}(\varrho_u, \varrho_s)$. 

The calculations required in (\ref{eq:piwls}) can be done very efficiently by employing generalized linear array methods, \cite{Currie:2006} which is possible due to the regular grid of bins underlying $\mathbf{Y}_\ell$, $\mathbf{R}$ and $\mathbf{E}_\ell$.

Once the estimated coefficients $\widehat{\mathbf{A}}_\ell =[\hat{\alpha}_{lm}^{\ell}]$ are obtained, we can evaluate the estimated $\hat{\breve{\eta}}_\ell (u,s)$ at arbitrary points  $(\dot u, \dot s)$ in the range of the two bases:

\begin{equation}
\label{eq:hateta}
\hat{\breve\eta}_{\ell}(\dot u,\dot s) = \mathbf{B}_u(\dot u)\hat{\mathbf{A}}_\ell \mathbf{B}_s(\dot s)\tp = \sum_{l=1}^{c_u}\sum_{m=1}^{c_s} b_l^u(\dot u)b_m^s(\dot s)\hat\alpha_{lm}^{\ell}.
\end{equation}
The cause-specific hazards are obtained by exponentiation of the linear predictors (\ref{eq:hateta})
\begin{equation}\label{eq:hatlambda}
\hat{\breve\lambda}_{\ell}(\dot u,\dot s) = \exp\left\{\hat{\breve\eta}_{\ell}(\dot u,\dot s)\right \}.
\end{equation}

\subsection{From cause-specific hazards to cumulative incidence functions}
\label{subsect:cif}

To obtain the overall survival function and the two cause-specific cumulative incidence functions, we first have to obtain the cumulative cause-specific hazards. In the model with two time scales this can be done in either of the two specifications, see equation (\ref{eq:2Dspec}):
\begin{equation}
\label{eq:cumhaz}
\Lambda _\ell(t,s) =\int _0 ^s \, \lambda_\ell(t=u+v, v) \, dv
\text{\qquad or \qquad}
\breve \Lambda _\ell(u,s) =\int _0 ^s \, \breve{\lambda}_\ell(u, v) \, dv.
\end{equation}
The overall survival function correspondingly is 
\begin{equation}
\label{eq:surv}
S(t,s) = \exp\left \{ -\sum_{\ell=1}^{2} \Lambda _\ell (t,s) \right \}
\text{\qquad or \qquad}
\breve S(u,s) = \exp\left \{ -\sum_{\ell=1}^{2} \breve \Lambda _\ell (u,s) \right\} .
\end{equation}
The survival functions above give, in the specific example, the probability of not having died by age $t$ and time since diagnosis $s$, that is $S(t,s)$, which is equivalent to surviving more than $s$ time units after diagnosis, when the disease was diagnosed at age $u=t-s$, that is $\breve S(u,s)$.

Finally, the two cumulative incidence functions (CIFs) are 
\begin{equation}
\label{eq:cif}
\begin{aligned}
I_\ell (t,s) &= \int _0 ^s  \lambda_\ell (t=u+v, v)\, S(t=u+v, v) \, dv  \\
\text{or} \\
\breve I_\ell (u,s) &= \int _0 ^s \breve\lambda_\ell (u, v)\, \tilde S(u, v) \, dv
\end{aligned}
\end{equation}
and give the probability of dying from cancer ($\ell=1$) or because of other causes other ($\ell=2$) before age $t$ and time since diagnosis $s$, or respectively, by $s$ time units after diagnosis that was made at age $u=t-s$.

Integration in (\ref{eq:cumhaz}) and (\ref{eq:cif}) is over one dimension only and we determine all integrals by numerical quadrature. Since we can evaluate the estimated CSHs on an arbitrarily fine grid, see (\ref{eq:hateta}), approximation errors will be minimal, even with a simple rectangle rule, as long as we choose a narrow enough grid width $\Delta$.
Hence we have, for the $(u,s)$-specification,

\begin{equation}
\hat{\breve \Lambda}_\ell(u, s) \approx 
\sum_{k=0}^{K({s})} \hat{\breve \lambda}_\ell(u, k \Delta) \, \Delta ,
\label{eq:hatcumhaz}
\end{equation} 
where $K(s)$ is the largest integer such that $K(s) \Delta < s$. Similarly, for the CIFs we obtain
\begin{equation}
\label{eq:hatcif}
\hat{\breve I}_\ell(u, s) \approx 
\sum_{k=0}^{K({s})}\, \hat{\breve\lambda}_\ell(u, k\Delta) \hat{\breve S}(u, k\Delta) \Delta .
\end{equation}

\subsection{Standard errors}
\label{subsect:se}
The variance-covariance matrix of the coefficients $\hat{\boldsymbol{\alpha}}_\ell$, $\ell = 1,2$, is given by \citep{EilersMarx:2021}
\begin{equation}
\label{eq:varalphal}
\boldsymbol{\Sigma}_\ell = (\mathbf{B} \hat{\mathbf{W}}_\ell \mathbf{B}\tp + \mathbf{P}_\ell)^{-1}.
\end{equation}
From this, the standard errors of other quantities can be derived via the delta method, such as 
\begin{equation}
\label{eq:seetal}
 \text{s.e.}(\,\hat{\breve\eta}_\ell (\dot u, \dot s) \,) = 
 \left (\mathbf B (\dot u, \dot s) \,\boldsymbol{\Sigma}_\ell \, \mathbf B (\dot u, \dot s) \tp \right )  ^{1/2} ,
 \end{equation}
where $(\dot u, \dot s)$ is an arbitrary point within the range of the marginal $B$-spline bases.
$\mathbf B(\dot u, \dot s)= \mathbf B_s(\dot s) \otimes \mathbf B_u(\dot u) $ is the model matrix evaluated at this point. Therefrom, we obtain for the cause-specific hazards 
\begin{equation}
\label{eq:selambdal}
 \text{s.e.}(\,\hat{\breve\lambda}_\ell (\dot u, \dot s) \,) = 
 \text{s.e.}  (\, \exp\{\hat{\breve\eta}_\ell(\dot u, \dot s)\} \,) = 
 \hat{\breve\lambda}_\ell (\dot u, \dot s) \;
 \text{s.e.}(\,\hat{\breve\eta}_\ell (\dot u, \dot s)\,) .
 \end{equation}
To calculate the standard errors for the cumulative incidence function, we use a simulation approach based on the assumption of asymptotic normality for the regression parameters. \cite{Mandel:2013} This method is commonly applied to derive uncertainty measures for non-linear functions with complex derivatives, such as cumulative incidence functions or other metrics derived from estimated rates. \cite{vanDenHout:2019, Jackson:2016} 

We repeatedly draw $\hat{\boldsymbol{\alpha}}_\ell^*$ from an asymptotic multivariate normal distribution with mean $\hat{\boldsymbol{\alpha}}_\ell$ and variance-covariance matrix as specified in equation~(\ref{eq:varalphal}). By generating a large number of samples (e.g., 10,000) from this distribution, we can estimate the CIF for each sample. The standard errors of the CIFs are then calculated as the square root of the empirical variance of these estimates.

\subsection{Ungrouping a final interval by the penalised composite link model}
\label{subsect:pclm}

Age information in data on cancer incidence or mortality is often available in age-groups only. Frequently the highest ages are combined in an open last interval. Such data  grouping is done to prevent identification of the patients \cite{Rizzi:2018} or to provide a compact representation of the data.

Grouped data do not represent a problem as such, but in some occasions it is desirable to work with data that are disaggregated to a finer resolution. 
The hazard model with two time scales that was introduced in Section~\ref{subsect:estimation} employs data that are binned in relatively narrow intervals of same size, resulting in data on a regular grid. The gridded structure allows efficient computations by using GLAM algorithms.

The SEER data, which we will analyse in Section~\ref{sect:application}, provide age at diagnosis in single years up to age 89, but group all observations with higher age at diagnosis in a single last age-group~90+. 
In the analyses we define this interval as $[90, 100)$, assuming that only very few, if any, diagnoses will be made beyond.
Time since diagnosis is given in months for all individuals, also for those in this last age-interval, so only one time dimension is affected by coarse grouping. 

To preserve the gridded data structure we will disaggregate the final age-group using the penalized composite link model (PCLM), originally introduced by Eilers.\cite{Eilers:2007} A Bayesian version was presented in Lambert and Eilers \cite{LambertEilers:2009} and Rizzi et al.~\cite{Rizzi:2015} used the model for ungrouping ages at death in one dimension. Here we describe the bivariate extension proposed in Rizzi et al. \cite{Rizzi:2018}

When all individual ages at diagnosis were available in single years as well as the individual follow-up times (in months), we denoted the binned event counts by the matrices $\mathbf{Y}_\ell = [y_{jk}^\ell]$ of dimension $n_u \times n_s$, see the description in Section~\ref{subsect:estimation}.  
Here, $j=1, \dots, n_u$ indexes the same-sized bins over age at diagnosis, and $k=1, \dots, n_s$ the same-sized bins over time since diagnosis. Similarly, the at-risk times were collected in the matrix $\mathbf{R}$ of the same dimension. 
The observed counts $y_{jk}^\ell$, cross-classified by age at diagnosis and time since diagnosis (both binned), are realizations of a multinomial distribution and can equivalently be viewed as realizations from independent Poisson variates\cite{BishopFienbergHolland} $Y_{jk}^\ell$ with expected values $\E(Y_{jk}^\ell) = \gamma_{jk}^\ell$. 

When the final age interval is (much) wider, we actually observe counts $z_{j'k}^\ell$, where the first $(g-1)$ elements along the age-at-diagnosis dimension are unchanged so that 
$z_{j'k}^\ell = y_{j'k}^\ell$ for $j'=1, \ldots , (g-1)$ and for all $k$. The last  count 
$z_{gk}^\ell$, however, represents the sum of the counts in the final wider age-group: 
$z_{gk}^\ell = \sum_{j=g} ^{n_u} y_{jk}^\ell$. Hence the matrices of event counts now are $\mathbf{Z}_\ell =[z_{j'k}^\ell ]$ of dimension $g\times n_s$. 

This summation carries over to the Poisson representation of the $Z_{j'k}^\ell$ with means $\E(Z_{j'k}^\ell) = \psi_{j'k}^\ell$, so that we have (in matrix notation) 

\begin{equation}
    \mathbf{\Psi}^\ell = \begin{bmatrix}
        \psi_{11}^\ell& \cdots &  \psi_{1n_s}^\ell\\
        \vdots& & \vdots \\
        \psi_{g1}^\ell&\cdots & \psi_{g n_s}^\ell\\
    \end{bmatrix}
\qquad
     \mathbf{\Gamma}^\ell = \begin{bmatrix}
        \gamma_{11}^\ell& \cdots &  \gamma_{1n_s}^\ell\\
        \vdots& & \vdots \\
        \gamma_{n_u 1}^\ell&\cdots & \gamma_{n_u n_s}^\ell\\
    \end{bmatrix}
\quad\text{and}\quad
     \mathbf{\Psi}^\ell = \mathbf{C}_u  \mathbf{\Gamma}^\ell
\end{equation}
with $\mathbf{C}_u$ being the marginal $g \times n_u$ composition matrix

\begin{equation}\label{eq:Cmat}
\mathbf{C}_u = \begin{bmatrix}
				~1 & \cdots &0 & \rvline & 0 & \cdots &0~\\
                ~\vdots & \ddots &\vdots & \rvline & \vdots&  &\vdots~\\
                ~0 & \cdots &1 & \rvline & 0& \cdots & 0~\\
			      \cmidrule[0.1pt](lr){1-7} %\hline 
				~0 &\cdots & 0& \rvline & 1& \cdots & 1~\\
			\end{bmatrix} \, .
\end{equation}
The top left sub-matrix is a $(g-1)$-dimensional identity matrix producing the unchanged values $\psi_{j'k}^\ell = \gamma_{j'k}^\ell$, the bottom row produces the $\psi_{gk}^\ell = \sum_{j=g}^ {n_u} \gamma_{jk}^\ell$ from the final wider age interval, for all $k=1, \ldots, n_s$.\footnote{A second composition matrix $C_s$ could constitute grouping along the second dimension, but since there is no such grouping in our case, $C_s$ reduces trivially to a $n_s$-dimensional identity matrix.}

Now the bivariate PCLM is employed to estimate the elements in $\mathbf{\Gamma} ^ \ell$ (on the finer and regular grid) from the observed counts $\mathbf{Z} _\ell$, the means of which are $ \mathbf{\Psi}^\ell = \mathbf{C}_u  \mathbf{\Gamma}^\ell$. The estimates in the rows $g$ to $n_u$ of $\widehat{\mathbf{\Gamma}}^\ell$ will then serve as the ungrouped event counts in the estimation of the cause-specific hazards (as described in Section~\ref{subsect:estimation}). 
Obviously, this problem is overparametrized with $g \, n_s$ data points for $n_u n_s$ parameters (and $g < n_u$). However, if we assume that the underlying distribution is smooth, we can express the 
$\mathbf{\Gamma}^\ell$ as linear combination of tensor products of $B$-splines and impose a  roughness penalty on the coefficients, as outlined in Section~\ref{subsect:estimation}. This penalization allows to estimate the elements in $\mathbf{\Gamma}^\ell$  from the fewer observed counts in $\mathbf{Z} _\ell$.

The PCLM for the observed counts ultimately is 
\begin{equation}
\label{eq:pclm}
\boldsymbol{z}^\ell \sim \text{Poisson}(\boldsymbol{\psi}^\ell)\quad\text{with}\quad
\boldsymbol{\psi}^\ell = \mathbf{C} \cdot \boldsymbol{\gamma}^\ell = 
\mathbf{C} \cdot \exp(\mathbf{B}\boldsymbol{\theta}^\ell) ,
\end{equation}
 where the vectors $\boldsymbol{z}^\ell$, $\boldsymbol{\psi}^\ell$ and $\boldsymbol{\gamma}^\ell$ contain the elements of the corresponding matrices. $\mathbf{C} = \mathbf{C}_s \otimes \mathbf{C}_u$ is the composition matrix with, in our case, an $n_s$-dimensional identity matrix 
 $\mathbf{C}_s$ and $\mathbf{C}_u$ as in equation~(\ref{eq:Cmat}).
 The regression matrix $\mathbf{B}$ contains the tensor products of two marginal $B$-spline bases, one for each time axis: $\mathbf{B}  = \mathbf{B}_s \otimes \mathbf{B}_u $, 
where $\mathbf{B}_u$ is of dimension $n_u \times c_u$ and $\mathbf{B}_s$ is of dimension $n_s \times c_s$. The vector $\boldsymbol{\theta}^\ell = (\theta_1^\ell, \dots, \theta_{c_uc_s}^\ell)\tp$ contains the $B$-spline coefficients. (The number of coefficients $c_u$ and $c_s$ in the $B$-spline representation can be different from the one for the CSHs, see equation~(\ref{eq:eta}), however, for notational parsimony we chose the same dimension here.).
The coefficients $\boldsymbol{\theta}^\ell$ are penalized by the penalty matrix 
\begin{equation}
\label{eq:Pstar}    
\mathbf{P}^\star= \phi_u (\mathbf{I}_{c_s} \otimes \mathbf{D}_u\tp \mathbf{D}_u) + \phi_s (\mathbf{D}_s\tp \mathbf{D}_s \otimes \mathbf{I}_{c_u}) \, ,
\end{equation}
where $\phi_u$ and $\phi_s$ are two smoothing parameters, $\mathbf{D}_u$ and $\mathbf{D}_s$ are difference matrices, usually of second order, and $\mathbf{I}_{c_u}$ and $\mathbf{I}_{c_s}$ are identity matrices of dimension $c_u$ and $c_s$ respectively. 

The penalized deviance of model (\ref{eq:pclm}), which has to be minimized, is
\begin{equation}
\label{eq:pdevpclm}
\text{dev}(\boldsymbol{\psi}^\ell, \mathbf{z}^\ell) + \text{pen}^\star = 2 \sum_{i = 1}^{n_g} \sum_{k = 1}^{n_s} \left ( z_{ik}^\ell \log(z_{ij}^\ell/\psi_{jk}^\ell) - (z_{ij}^\ell - \psi_{ik}^\ell) \right ) + \text{pen}^\star,
\end{equation}
where 
\begin{equation}
\label{eq:penstar}
\text{pen}^\star (\phi_i, \phi_s) = \phi_u \, || \mathbf{D}_u \boldsymbol{\psi}^\ell||^2_{\text{\tiny F}} + \phi_s \, ||\boldsymbol{\psi}^\ell \mathbf{D}_s \tp||_{\text{\tiny F}}^2 .
\end{equation}
The linearized score equations again can be solved by an IWLS algorithm, for fixed values of $\phi_u$ and $\phi_s$. 
To select the optimal values of the smoothing parameters $\phi_u$ and $\phi_s$ we minimize $\text{AIC}(\phi_u ,\phi_s)$.

Once the PCLM is estimated, the ungrouped values replace the event counts in the final age interval so that in $\mathbf{Y}_\ell$ we use $y_{jk}^\ell = \hat\gamma_{jk}^\ell$ for $j=g,\ldots , n_u$ and all $k= 1,\ldots, n_s$.

To estimate the cause-specific hazards not only the event counts but also the at-risk times in the final age-interval have to be expanded onto the finer regular grid. For that purpose, the number of individuals at risk in the last age-group 90+ at each follow up time, is ungrouped following the strategy presented above, to obtain an estimate of the number of individuals at ages [90-100) entering each bin of the follow up. From these, an estimate of the exposure times corresponding to ages  [90-100) is obtained by assuming that each individual who exits a bin (due to event or end of follow-up) contributes half of the total bin's length and each individual who survives the bin contributes exposure time for the full length of the bin.

We discuss how we used the model described here to ungroup ages at diagnosis in our data in sub-section~\ref{subsect:SEERungroup}.

%\newpage

\section{Mortality of women with breast cancer}
\label{sect:application}

\subsection{The SEER data}
\label{subsect:data}

The Surveillance, Epidemiology and End Results (SEER) program from the National Cancer Institute collects and publishes data on cancer incidence and survival covering about 48\% of the population of the USA. \cite{SEER} In this study we use data from the incidence SEER-17 database (November 2022 submission \cite{SEERdata}), that includes all cancer diagnoses up to 2020. We accessed the registers on 09/08/2023 and used the case listing functionality of SEER*Stat \cite{SEERsoftware} to extract all cases of malignant breast cancer (primary site codes C50.0-C50.9) diagnosed to women between 2010 and 2015, with maximum follow-up to the end of 2020. 

We focus on postmenopausal women and hence selected only women who were at least 50 years old when the cancer was diagnosed. Year of diagnosis was restricted to 2010--2015 to limit potential period effects, such as improvements in diagnosis and treatment efficacy that may affect survival probabilities. The potential maximum length of follow-up was several years for all women in the sample (although of different length, depending on the specific year of diagnosis).

Only individuals for whom the cancer was the first malignant tumor ever recorded in the SEER registry were included in the analysis. Furthermore, only cases for which the cancer was confirmed microscopically or by a positive test/marker study were included, while cases where the cancer was detected only by autopsy or only reported on the death certificate were not considered. 
 
Mortality of breast cancer patients is known to depend on race/ethnicity, \citep{Campbell:2002} independent of socioeconomic factors. SEER provides a variable that combines information on race and ethnicity and we distinguish between white women, black women and women of race/ethnicity other than white or black in our analysis. 
Also the cancer sub-type influences the survival chances, \citep{Brandt:2015, Jayasekara:2019} and the sub-types are categorized according to their hormone receptor status (HR+ or HR-) and to the status of the human epidermal growth factor 2 receptor (HER2+ or HER2-) \cite{Howlader:2018}. Different combinations of the receptor statuses can be grouped further, and we follow the distinction between the Luminal~A sub-type (HR+ and HER2-) and all other sub-types of breast cancers. Luminal~A is the most common and least aggressive sub-type of breast cancer and it is linked to higher survival probabilities. \cite{Howlader:2018}

Cases for which no information on the cancer sub-type was available (7.8\%) or where race/ethnicity information was unknown (0.4\%) were not included in the analysis.
SEER also provides information on whether chemotherapy was performed, but does not distinguish between cases where chemotherapy was not given and those where this information is missing. The final size of the dataset is 202,242 women. Table~\ref{tab:sample} gives a summary breakdown by race/ethnicity, cancer sub-type and chemotherapy status. 

\begin{table}[]
\begin{tabular}{l|ll|ll|ll}
Race/ethnicity    			& \multicolumn{2}{c|}{White} & \multicolumn{2}{c|}{Black} & \multicolumn{2}{c}{Other race/ethnicity} \\
\hline
Cancer subtype    			& Luminal A     & ~~Other     & Luminal A     & ~~Other     & Luminal A     &~~Other     \\
\hline
Chemotherapy      & \multicolumn{1}{r}{30,245}       & \multicolumn{1}{r|}{25,878}    & \multicolumn{1}{r}{4,259 }        & \multicolumn{1}{r|}{5,372}       &              \multicolumn{1}{r}{3,580}       & \multicolumn{1}{r}{3,498}       \\
Unknown/No Chemo     & \multicolumn{1}{r}{95,424 }        & \multicolumn{1}{r|}{11,701}    &\multicolumn{1}{r}{8,704}         & \multicolumn{1}{r|}{2,016}      & \multicolumn{1}{r}{10,776}         & \multicolumn{1}{r}{1,498}  \\
\hline
\end{tabular}
\caption{\label{tab:sample} Sample size in each group identified by race/ethnicity, breast cancer subtype and chemotherapy status.}
\end{table}

The key variables for our analysis are age at diagnosis, time since diagnosis, vital status at the end of the follow-up and, in case of death, whether death was attributable to the cancer or to any other cause of death.
Time since diagnosis is available in the registry as survival months, that is the number of months from diagnosis to either death or right-censoring. 

In the raw data, end of follow-up is fixed at the 31st of December 2020. However, since the Covid-19 pandemic in 2020 likely had a noticeable impact on cancer patients, we decided to limit the follow-up in our analyses to the end of 2019.
For individuals who died before the end of 2019 no changes were required. For those individuals who were alive at the end of 2020 we subtracted 12 months of survival time and they were still right-censored. Finally,  individuals that exited the risk set during the year 2020 were right-censored at the end of 2019 and their survival times adjusted by six months. 

The different combinations of race/ethnicity, cancer sub-type and  chemotherapy indicator define 12 groups of interest for our analysis. The sizes of the subgroups allow separate analyses so we estimate all models separately and compare the results among different groups. In the following we will present a subset of the analyses and refer to the supplementary material for more results.

\subsection{Ungrouping ages at diagnosis with the PCLM}
\label{subsect:SEERungroup}
Age at diagnosis is provided in the SEER registry as single years of age, up to age 89, and a last open-ended interval 90+.
\begin{figure}[tbp]%
\centering
	\includegraphics[width=.5\columnwidth]{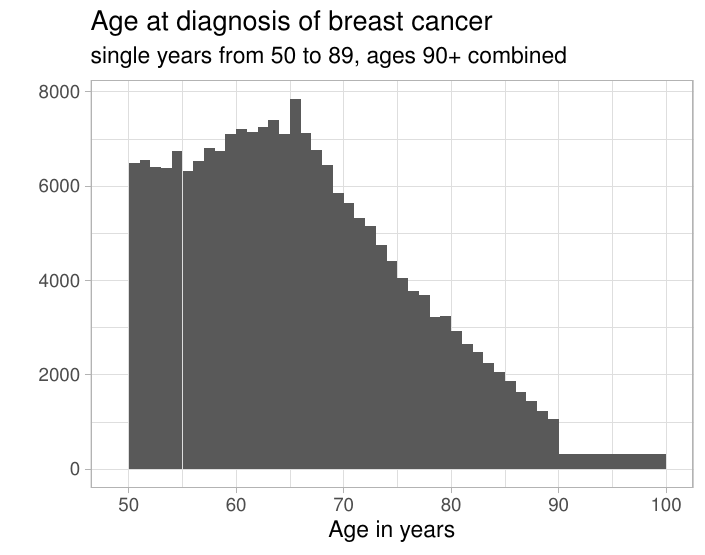}%
\caption{Distribution of age at diagnosis of breast cancer from the SEER data.}%
\label{fig:agediagn}%
\end{figure}
The distribution of age at diagnosis is shown in Figure~\ref{fig:agediagn}, for the whole sample. Two features are clearly visible: first, the grouping of ages at diagnosis greater or equal to 90, and second the small peak of the distribution at age 65. The latter feature can be explained by the enrollment in the Medicare program, which starts at age 65 and has been found to be connected to an increase in diagnoses of the most common cancers in the US. \cite{Patel:2021}
Of the twelve subgroups (see Table~\ref{tab:sample}) only eleven had observations with diagnoses made at age~90 or older, so the PCLM had only been applied in those subgroups.

Before we ungroup the ages at diagnosis in the interval $[90, 100)$, we first need to define the bins for the finer grid from which the cause-specific hazards will be estimated by the methods described in Section~\ref{subsect:estimation}. For all models we chose bins of length $h_u =1$ year for the age at diagnosis $u$ and $h_s=0.5$ year for the time since diagnosis $s$, which is given in months. This leads to 50 bins along the $u$-axis and $21$ along the $s$-axis.

The cause-specific event counts, and consequently the number of individuals at risk, 
for ages at diagnosis in the final wider interval are ungrouped using the PCLM as described in Section~\ref{subsect:pclm} and  the model specification is the same for all groups. 
The marginal $B$-spline matrices entering in 
$\mathbf{B}=\mathbf{B}_s \otimes\mathbf{B}_u$ in equation (\ref{eq:pclm}) are of dimension $50 \times 16$ for $\mathbf{B}_u$ and of dimension $21\times 10$ for $\mathbf{B}_s$. The composition matrices are built as explained in (\ref{eq:Cmat}), $C_u$ is of dimension $41 \times 50$ and $C_s$ is an identity matrix of dimension $21 \times 21$. 
The number of parameters in $\boldsymbol{\theta}^\ell$ to be estimated in the PCLM are $16\cdot 10 = 160$.
Finally, the optimal smoothing parameters $\hat\phi_u$ and $\hat\phi_s$ are selected, separately for each group, as those that produce the model with minimum AIC. This is done by grid search using values of $\log_{10}(\phi_u)$ and $\log_{10}(\phi_s)$ ranging from $-1$ to $2$. 

\begin{figure}
    \centering
    \includegraphics[width=.45\textwidth]{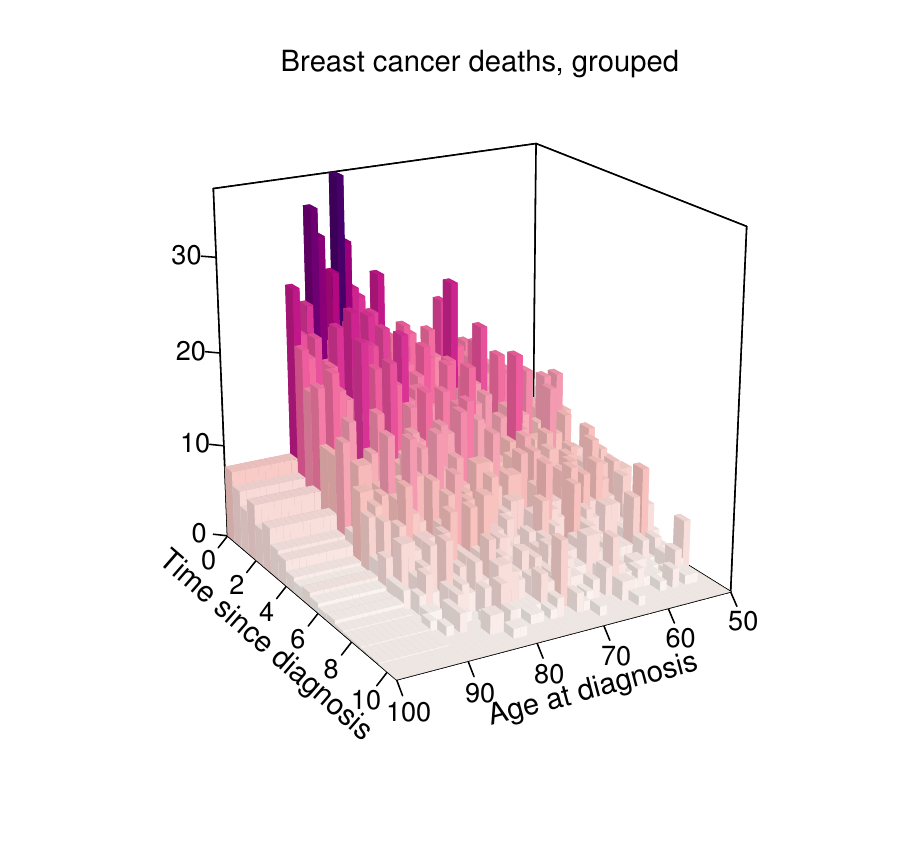}
    \includegraphics[width=.45\textwidth]{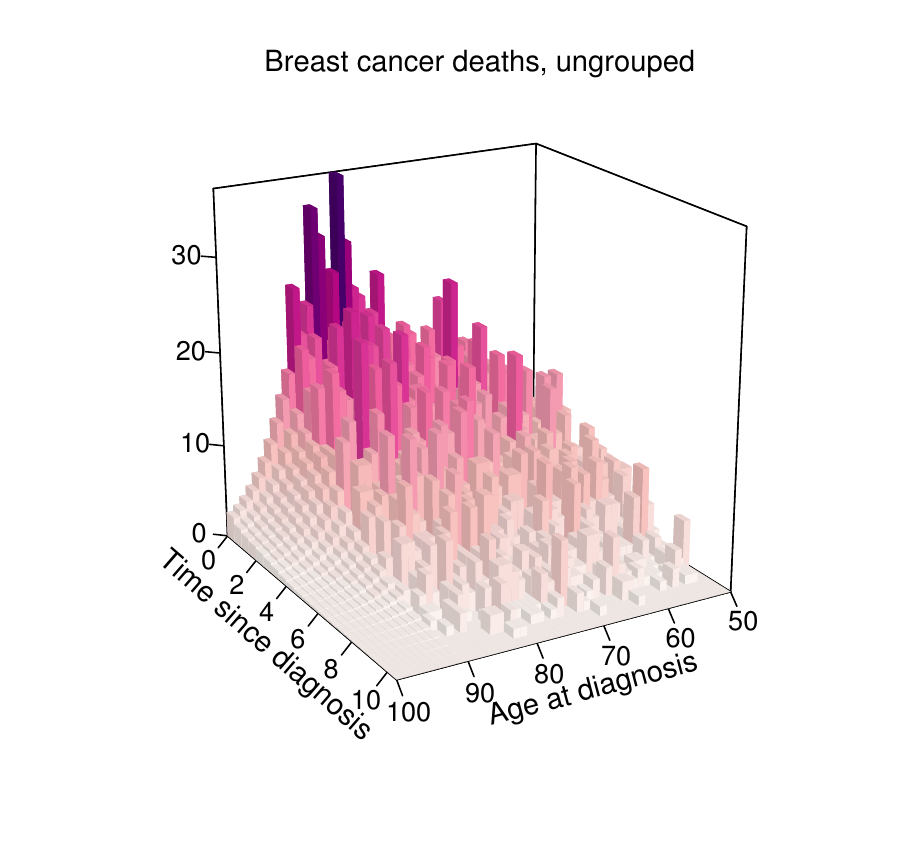}   
    \caption{Left: Histogram of deaths due to breast cancer over age at diagnosis (with ages $\geq 90$ grouped) and time since diagnosis. Right: Data ungrouped by the PCLM.}
    \label{fig:BCdeaths}
\end{figure}

Figure~\ref{fig:BCdeaths} illustrates this approach for white women with luminal A subtype and no chemotherapy. The raw counts of breast cancer deaths are plotted in the left panel, the last age-interval is 10 years wide and corresponds to ages $[90, 100)$. In the right panel, the counts that were ungrouped to single ages are plotted over ten bins of width 1.

After having obtained estimates for the ungrouped event counts and exposure times, we replace the grouped data in $[90, 100)$ with the estimated quantities and we use the so obtained matrices $\mathbf{Y}_\ell$ and $\mathbf{R}$ to estimate the cause-specific hazards and derived quantities as described in the next section. %~\ref{subsect:SEERcsh}.

\subsection{Estimating the cause-specific hazards, cumulative incidence functions and overall survival probabilities}
\label{subsect:SEERcsh}
\begin{figure}[h!]
    \centering
    \includegraphics[width=0.7\textwidth]{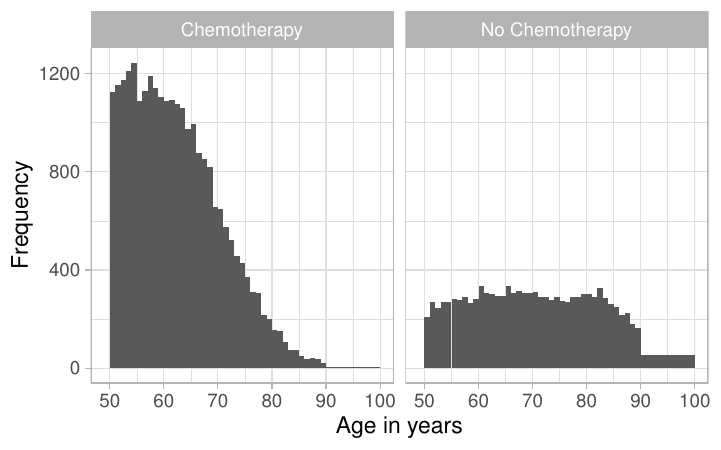}
    \caption{White women with cancer type other than Luminal A, age at diagnosis.}
    \label{fig:histAgeDiag}
\end{figure}

To illustrate the approach we present here the results for white women diagnosed with breast cancer of type other than Luminal A,  both for patients who received chemotherapy and who did not. The distribution of the ages at diagnosis is given in Figure~\ref{fig:histAgeDiag}. Noticeable grouping of ages~90+ is discernible for women that received no chemotherapy.

The matrices of event counts $\mathbf{Y}_\ell$ ($\ell=1$ for death due to breast cancer, $\ell=2$ for other causes) and the matrix $\mathbf{R}$ of exposure times are all of dimension $n_u \times n_s= 50\times 21$. 
To estimate the cause-specific hazards we chose 16 cubic $B$-splines on the $u$-axis and 10 cubic $B$-splines on the $s$-axis, see equation (\ref{eq:eta}). Penalty order was $d=2$. The optimal smoothing parameters $\varrho_u$ and $\varrho_s$ were determined by numerical optimization of BIC, see equation~(\ref{eq:AICBIC}). All computations were performed using the R-package \texttt{TwoTimeScales}. \cite{2TS} 

\begin{figure}[h!]
    \vspace*{2ex}
    \centerline{\sffamily \large No Chemotherapy\qquad\quad}
    \includegraphics[width=\textwidth]{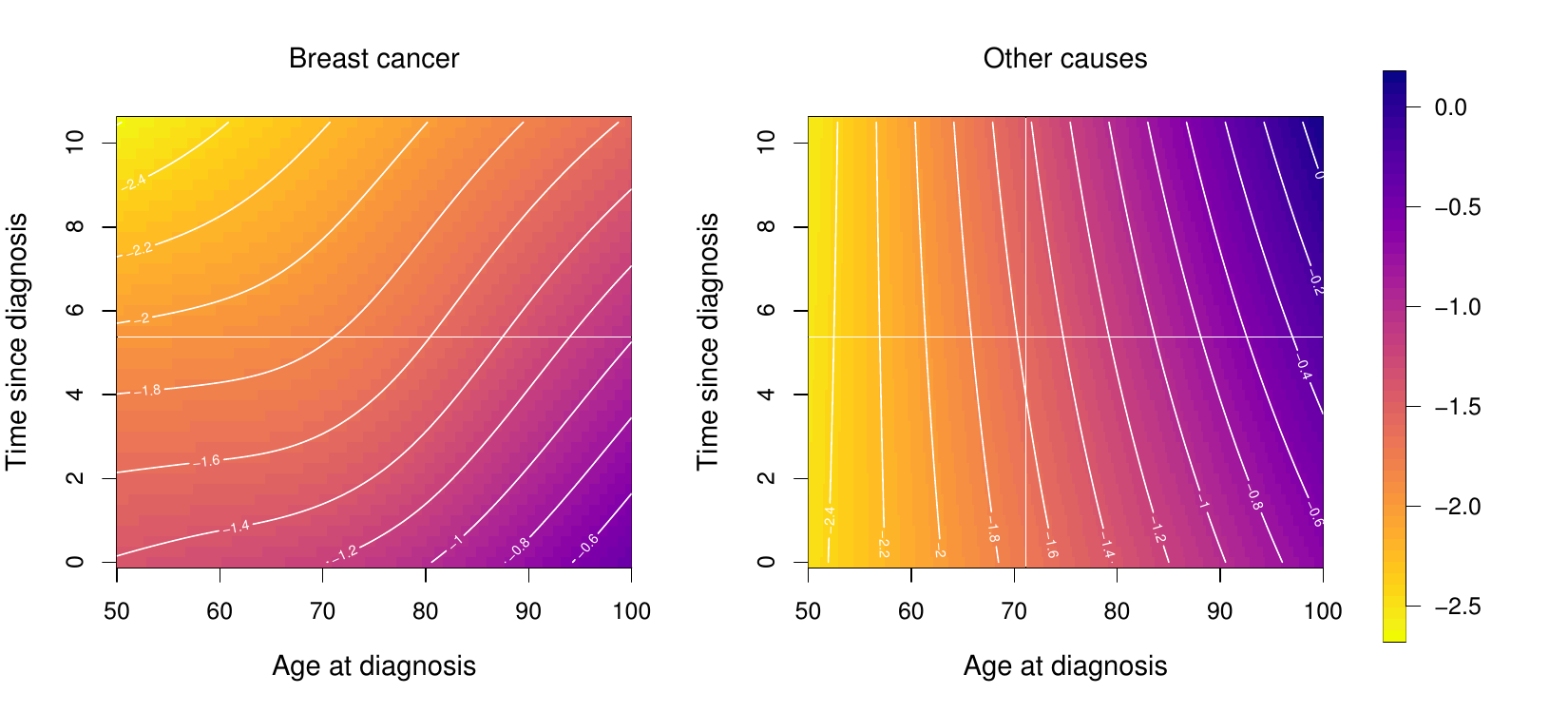}
 \smallskip

\centerline{\large\sffamily Chemotherapy\qquad~~}
    \includegraphics[width=\textwidth]{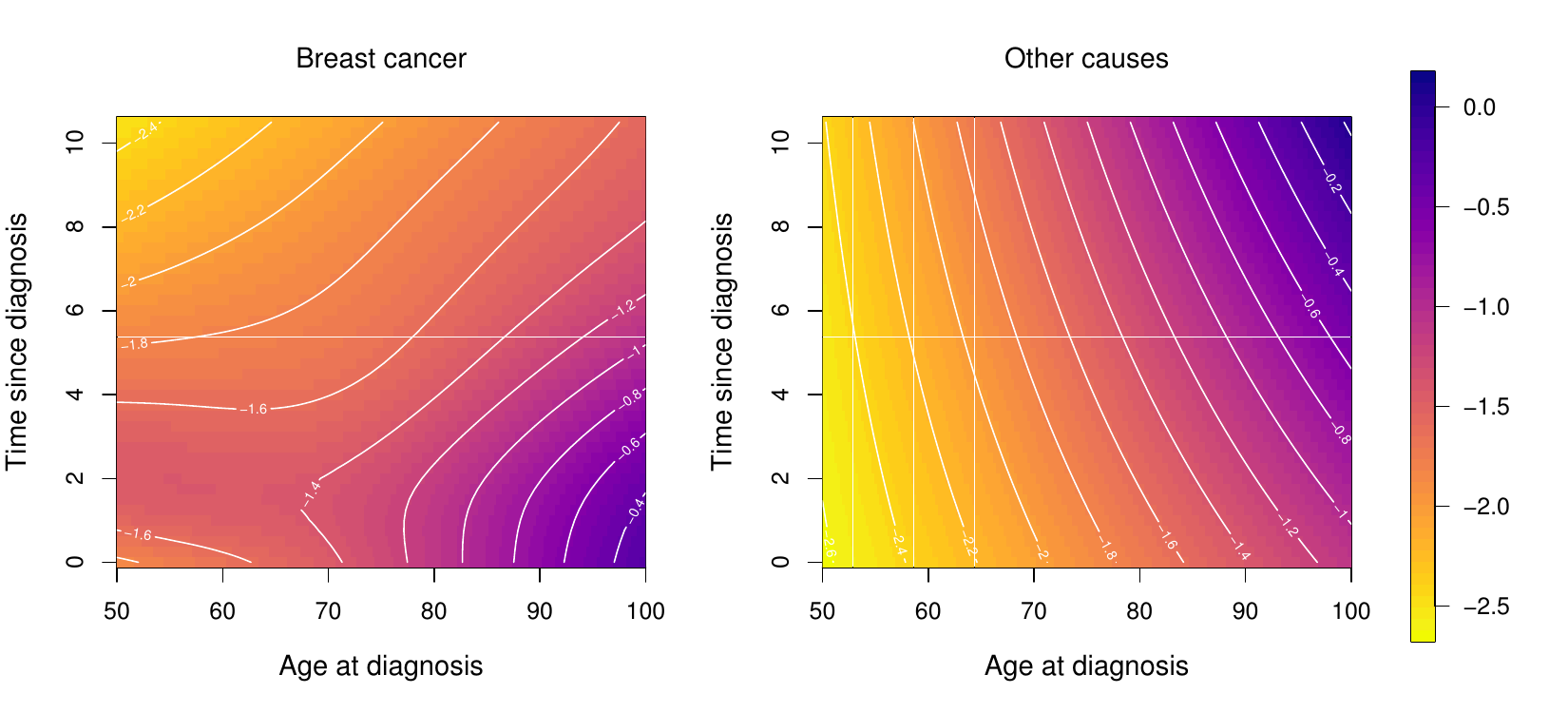}   
   
    \caption{$\text{Log}_{10}$ of cause-specific hazards for white women, other cancer types (non-Luminal A) who received chemotherapy (bottom row) and who did not (top row). Effective dimensions are given in Table~\ref{tab:EDs}. 
%    \textcolor{red}{To do: Adjust transparency for empty parts of risk sets}
}
    \label{fig:CSHwhiteOther} 
\end{figure}

Table~\ref{tab:EDs} summarizes the optimal smoothing parameters and the resulting effective dimensions for the two groups and the two causes of death, respectively. 
The estimated cause-specific hazards, on log$_{10}$-scale, are shown in Figure~\ref{fig:CSHwhiteOther}. 
\begin{table}
    \centering
\footnotesize
\begin{tabular}{lrrr c rrr}
\hline
&&&&&&&\\[-1mm]
     & \multicolumn{3}{c}{\small Death due to breast cancer}&&\multicolumn{3}{c}{\small Other causes of death}\\[1ex]
     
    & $\log_{10}\varrho_u$ & $\log_{10}\varrho_s$ & ~~~~ED && $\log_{10}\varrho_u$ & $\log_{10}\varrho_s$ & ~~~~ED\\
\cmidrule(lr){2-4}  \cmidrule(lr){6-8}
No chemotherapy      & 2.5 & 1.7 & 7.2 &&  7.5 &  6.9&  4.0\\
Chemotherapy         & 2.5 & 0.2 & 12.3&&  3.3 &  6.5 &  4.2\\[3.5mm]
\hline
\end{tabular}
\caption{White women with other subtypes: Optimal smoothing parameters (minimal BIC) and resulting effective dimensions (ED) for cause-specific hazards.} 
\label{tab:EDs}
\end{table}

For women who did not receive chemotherapy the hazard of dying of breast cancer declines with time since diagnosis and it is generally higher if it was diagnosed at a higher age. If women did receive chemotherapy then, for age at diagnosis up to about age 75, the cause-specific hazard for death due to breast cancer first increases and peaks about two years after diagnosis and declines thereafter. The different complexity of the hazard patterns is also reflected in the effective dimensions of the estimated hazard surfaces. 
The hazard of dying from other causes is, for both treatment groups, much smoother and reveals the well known exponential increase (Gompertz hazard) for ages~50+, however, the hazard level is lower for patients receiving chemotherapy but the increase in mortality over time is stronger for them.

The penalty allows to extend the estimated hazards into areas that are not supported by data. For example, in the specific application presented here, we know that for women who received chemotherapy and were diagnosed between age 90 and 100 the longest observed time since diagnosis was $s=6.75$ years. Hence the top right area of the cause-specific hazards (above $s=6.75$ and for $90 \leq u < 100$) is extrapolated beyond the observed data. 

\begin{figure}[h!]
    \centering
    \includegraphics[width=\textwidth]{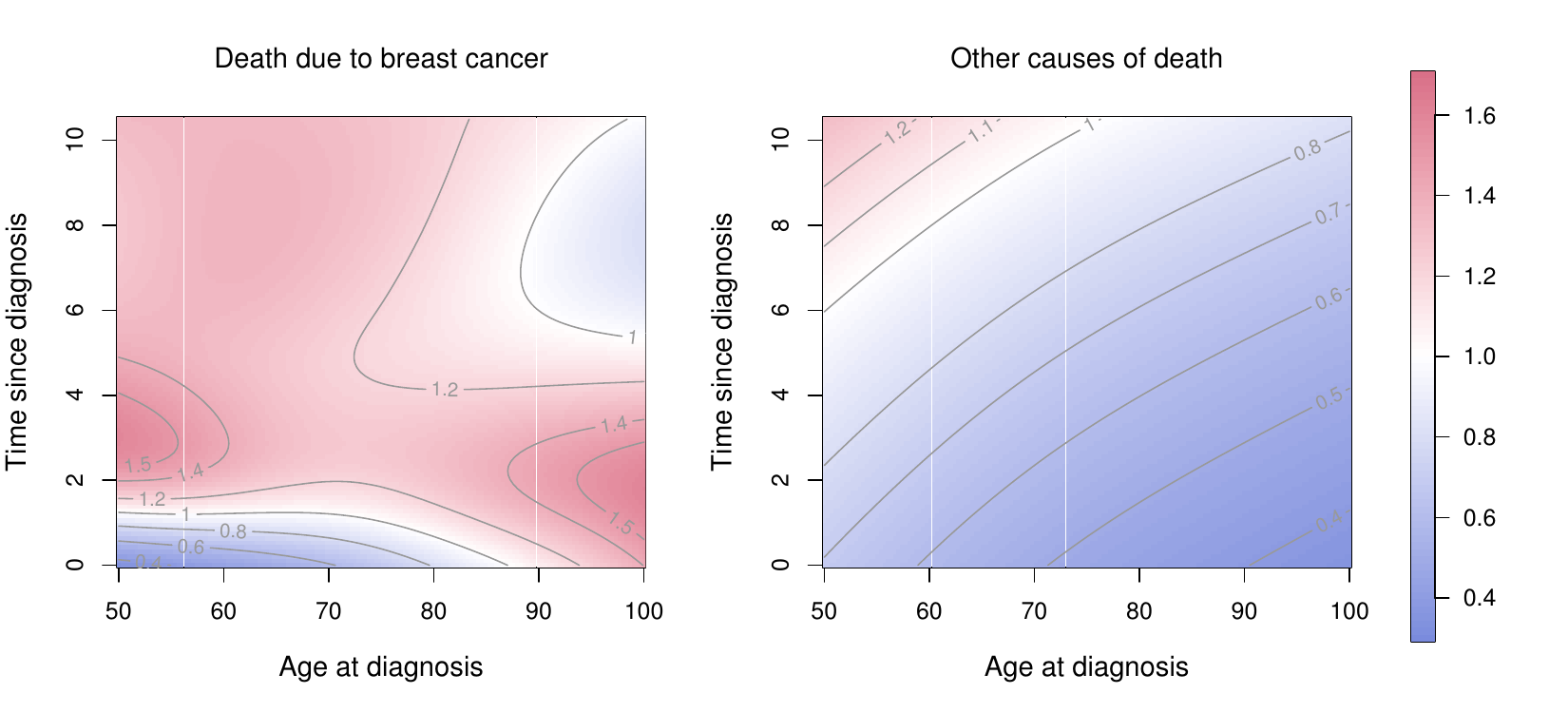}
    \caption{Ratio of the cause-specific hazards in Figure~\ref{fig:CSHwhiteOther}; chemotherapy versus no chemotherapy.}
    \label{fig:HRwhiteOther}
\end{figure}

It is already evident from Figure~\ref{fig:CSHwhiteOther} that application of chemotherapy leads to a more complex change in the hazards than a proportional shift, but to investigate the differences further Figure~\ref{fig:HRwhiteOther} shows the hazard ratios of the two cause-specific hazards (for women who received chemotherapy relative to those who did not).

\begin{figure}[h!]
    \centering
    \includegraphics[width=0.80\textwidth]{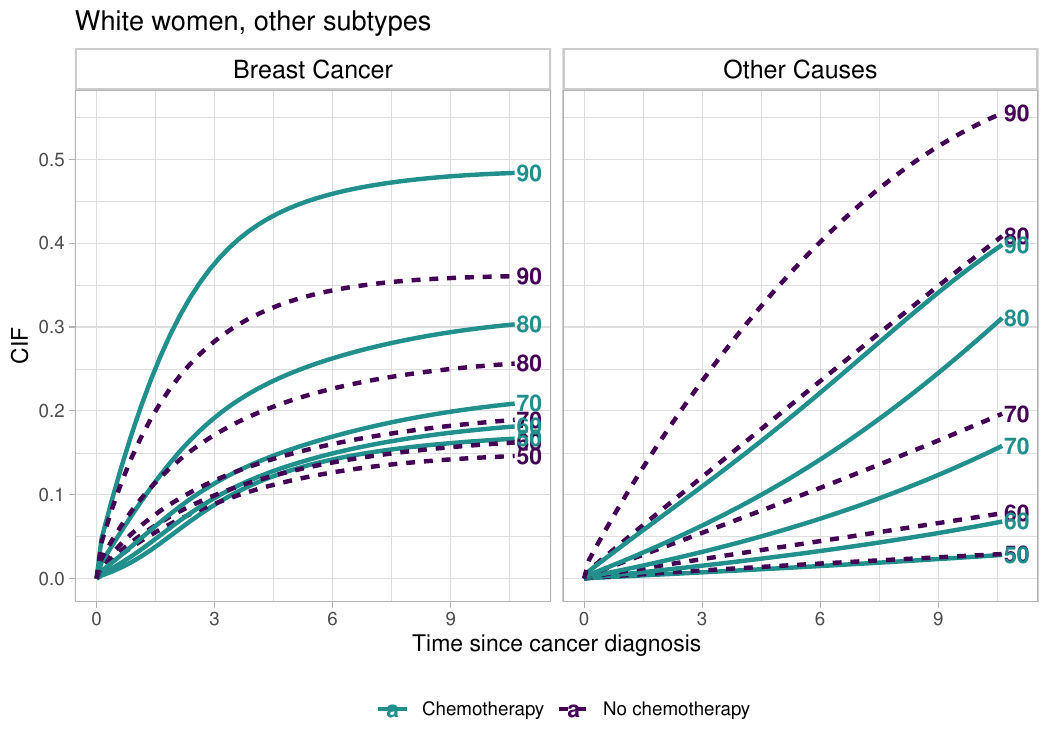}
    \caption{Cumulative incidence functions (CIF) for selected ages at diagnosis over time since diagnosis. The complete surfaces of the CIF can be found in the Appendix}
    \label{fig:CIFlinesWhiteOther}
\end{figure}

From the cause-specific hazards the overall survival function and the cumulative incidence functions can be derived, see Section~\ref{subsect:cif}. 
In Figure~\ref{fig:CIFlinesWhiteOther} we compare the cumulative incidence of death due to breast cancer and due to other causes for the treatment groups. The different lines correspond to different ages at diagnosis and result from transsecting the surfaces that are presented in the Appendix in Figure~\ref{fig:CIFwhiteOther} vertically. The associated SEs and the overall survival probabilities for the two treatment groups can also be found there.

The gap of the cumulative incidence between the treatment groups widens for ages at diagnosis above 70, but while for breast cancer deaths the group that received chemotherapy shows higher values of the CIF, for the other causes of death the order reverses. For younger ages at diagnosis and time since diagnosis up to about three years the CIFs reflect the unimodal shape of the cause-specific hazard in the chemotherapy group.

The two treatment groups are of quite different sizes (25,878 who received chemotherapy versus 11,701 who did not) and this, together with the distribution of observations across the $(u,s)$-plane, has an impact on the uncertainty of the estimates. For reasons of space standard errors for the cause-specific hazards are presented in the Supplementary Material. 

%\newpage
\section{Discussion}
\label{sect:discussion}
In this article we present a method to estimate a competing risks model where the cause-specific hazards and hence also the derived quantities in the model vary over two time scales. The model is estimated by two-dimensional $P$-splines and the only assumption made is that the hazards for the competing events vary smoothly over the two time scales. For estimation the event counts and the exposure times are binned in regular narrow two-dimensional intervals so that the well-known correspondence between hazard estimation and Poisson regression can be exploited. Computations can be performed very efficiently by applying GLAM algorithms. From the cause-specific hazards one can obtain overall survival probabilities and cause-specific incidence functions by simple numerical integration. As the estimated $P$-splines can be evaluated on arbitrarily fine grids the approximation error will be minor, even if simple integration techniques are used. The model is implemented in the R-package \texttt{TwoTimeScales}, which is available on GitHub. \cite{2TS} 

The GLAM approach rests upon data that are binned on a regular grid so that the array structure can be utilized in the computations. In practise we may encounter event times that are grouped in a wide interval for one or both time axes at the upper end of the time scale. The SEER data that we used for illustration of the approach in this paper provide such an example. To preserve the gridded structure of the data for estimating the cause-specific hazards we used the penalized composite link model to ungroup the final wide intervals. Again only smoothness of the underlying distributions is assumed and bivariate $P$-splines are employed. 
In this way the ungrouped data that enter the estimation procedure will be more smooth than the (unobserved) original data. Although the numbers in the final interval are rather low, concerns may arise that this could influence the smoothing parameters chosen by minimizing BIC in the estimation step. As a sensitivity analysis, we restricted the data to known ages at diagnosis, i.e., ages 50 to 89, to avoid the ungrouping step and then estimated the model from these data. The results of the comparison can be found in the Supplementary Material and demonstrate that the ungrouping by the PCLM and subsequent estimation alters the results only marginally.

In the application we considered race/ethnicity, tumor type and chemotherapy status and analysed subgroups separately which, in this case, was feasible due to the large sample size of the SEER registers. This allowed, on the one hand, to use the PCLM for ungrouping as this model operates on aggregated count data. It also allowed to identify deviations from proportional hazards for the two treatment groups that were particularly pronounced for the hazard of death due to breast cancer. Nevertheless, for more covariates and less comprehensive data sets hazard regression models will be relevant tools for analysis, also for cause-specific hazards that vary along two time scales. 
For exactly observed event times proportional hazards regression has been developed already \cite{Carollo:2023} and is also implemented in the R-package \texttt{TwoTimeScales}. \cite{2TS} The extension of this model to individual interval-censored data will be future research.

\newpage

% bibliography here

\newpage
\appendix
\section{Appendix}
\label{app:A}
\subsection{Further results of Section~\ref{sect:application}}
\label{app:SurvCIF}
\vskip 5ex

\begin{figure}[h!]
    \centering
    \centerline{\sffamily No Chemotherapy\qquad\quad}
  
    \includegraphics[width=0.9\textwidth]{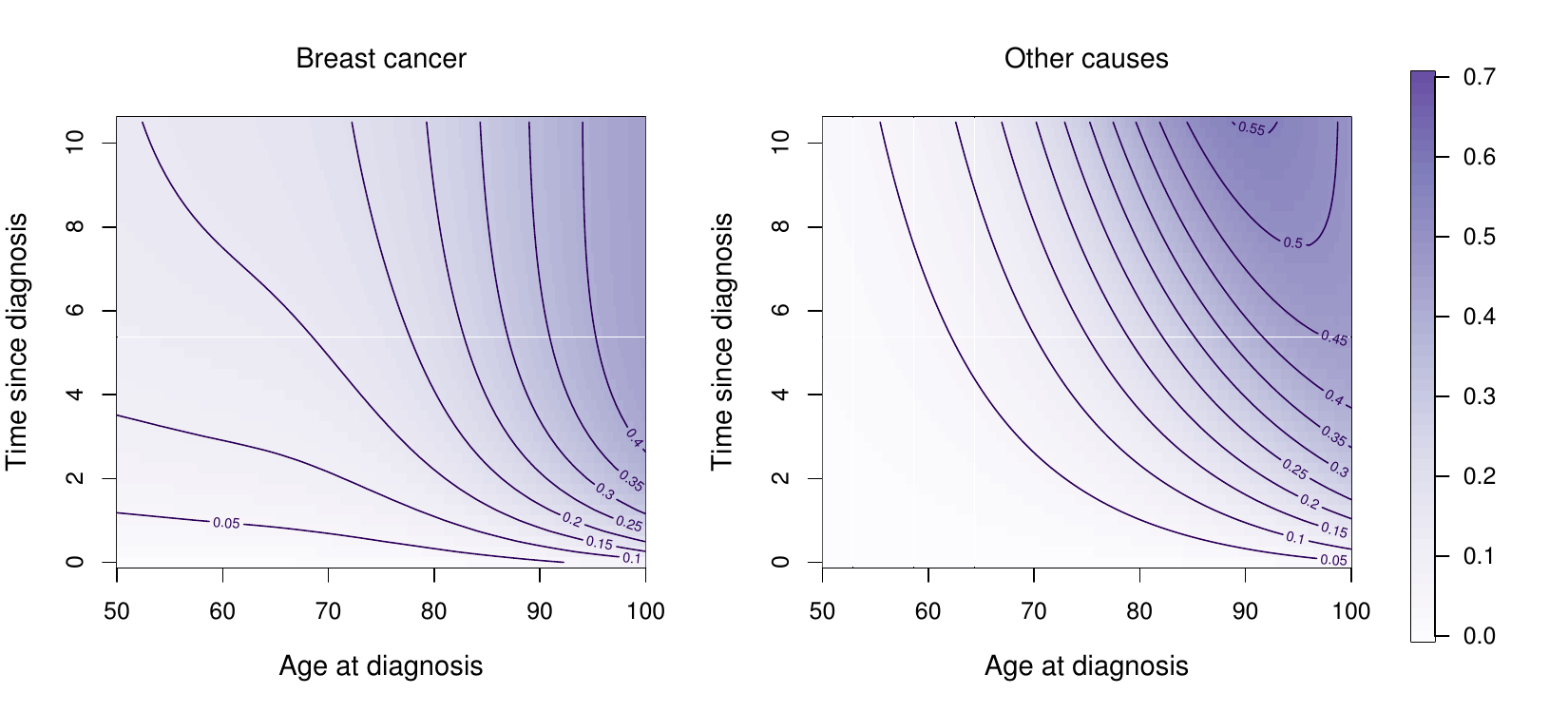} 
     \centerline{\sffamily Chemotherapy\qquad~~}
   
    \includegraphics[width=0.9\textwidth]{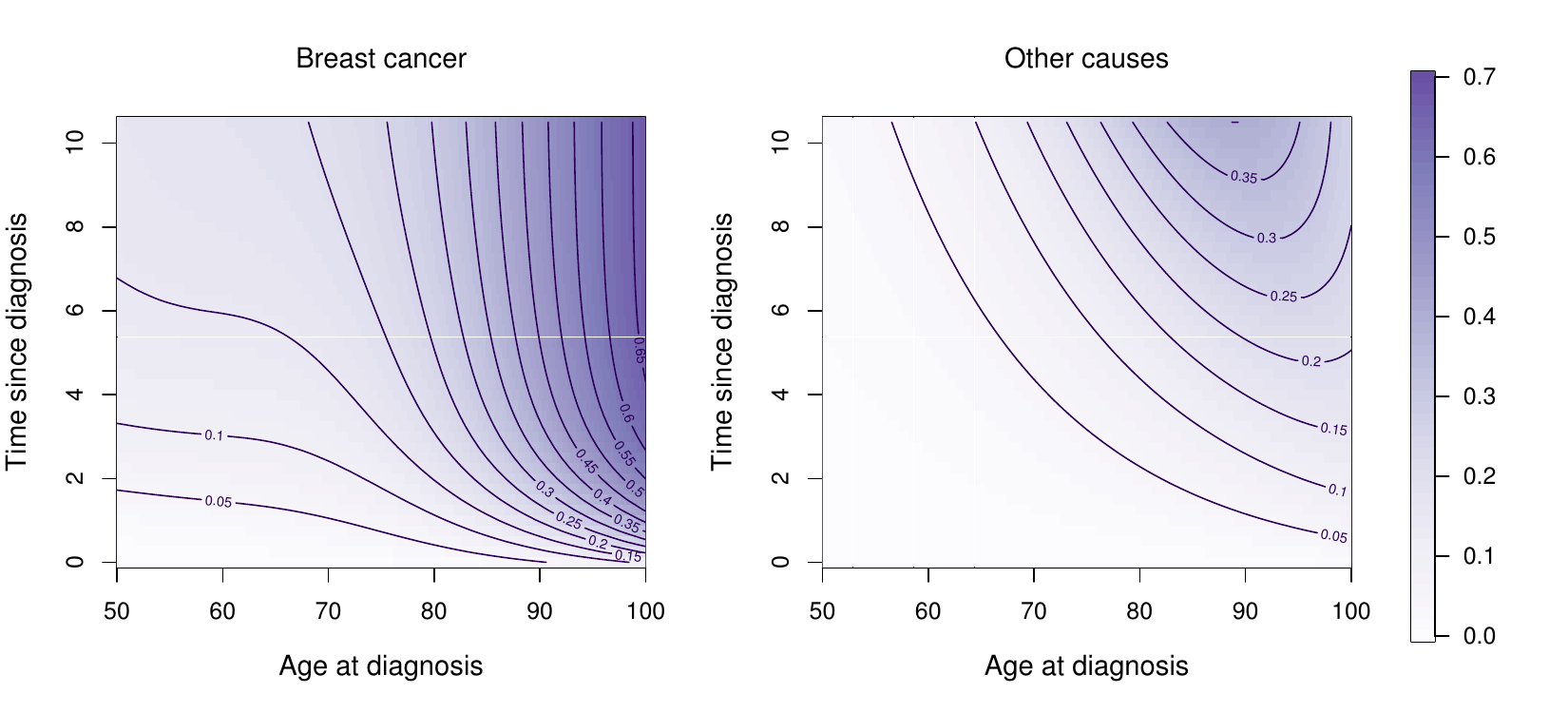} 
    \caption{Cumulative incidence functions for the cause-specific hazards displayed in Fig.~\ref{fig:CSHwhiteOther}, see also Fig.~\ref{fig:CIFlinesWhiteOther}. 
}
    \label{fig:CIFwhiteOther}
\end{figure}

\vskip 5ex

\begin{figure}[h!]
    \centering
    \centerline{\sffamily No Chemotherapy\qquad\quad}
  
    \includegraphics[width=0.9\textwidth]{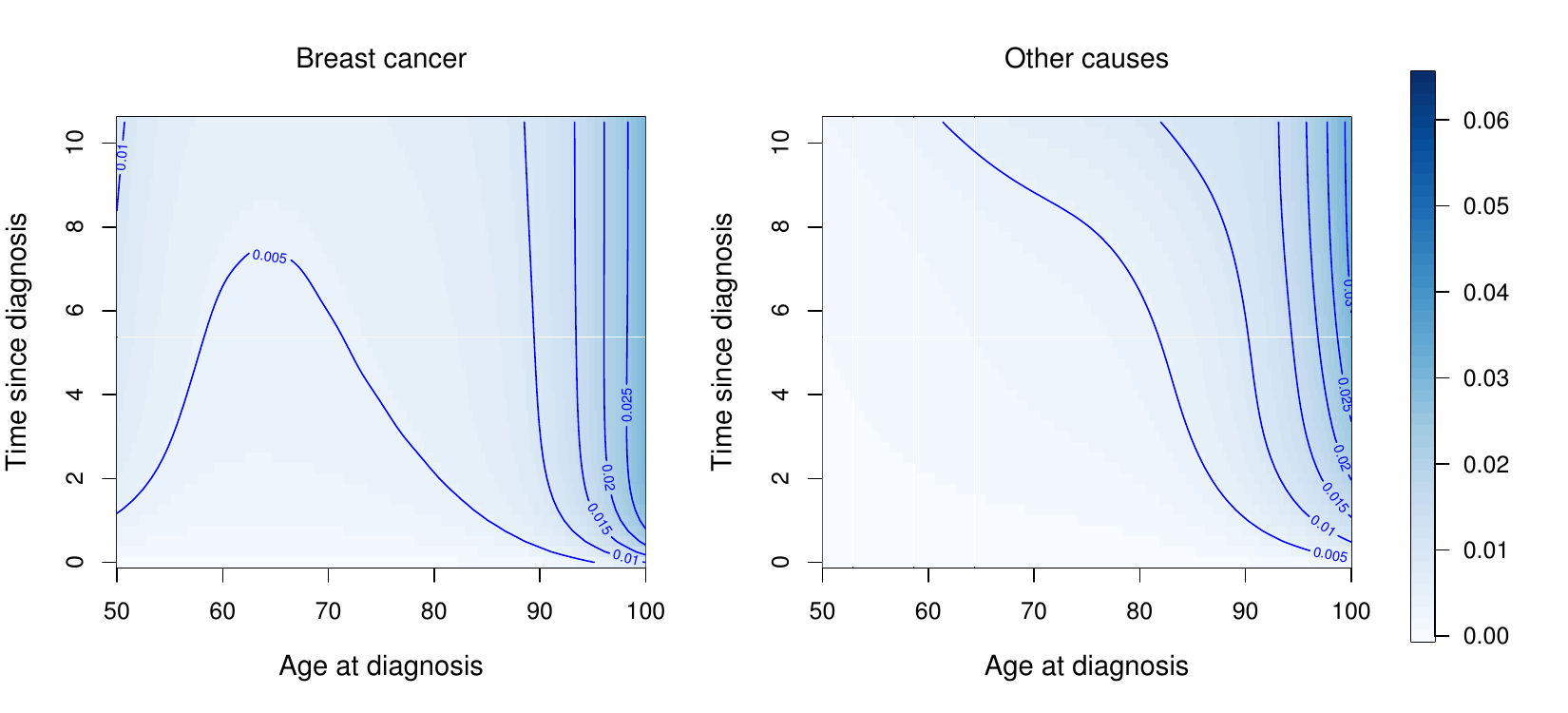} 
     \centerline{\sffamily Chemotherapy\qquad~~}
   
    \includegraphics[width=0.9\textwidth]{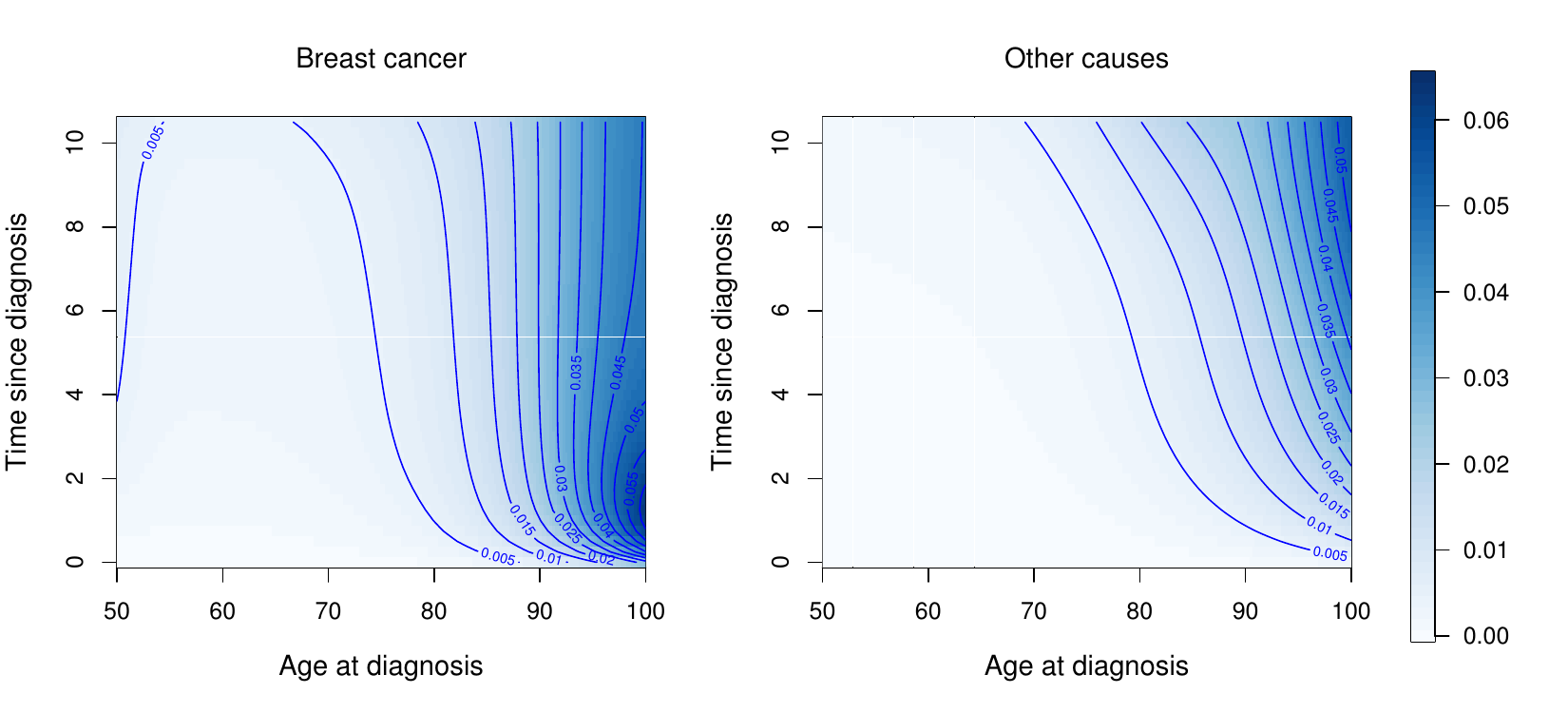} 
    \caption{Standard Errors of the cumulative incidence functions for the cause-specific hazards displayed in Fig.~\ref{fig:CIFwhiteOther}. 
}
    \label{fig:SEsCIFwhiteOther}
\end{figure}

\begin{figure}[ht]
    \centering
    \includegraphics[width=0.9\textwidth]{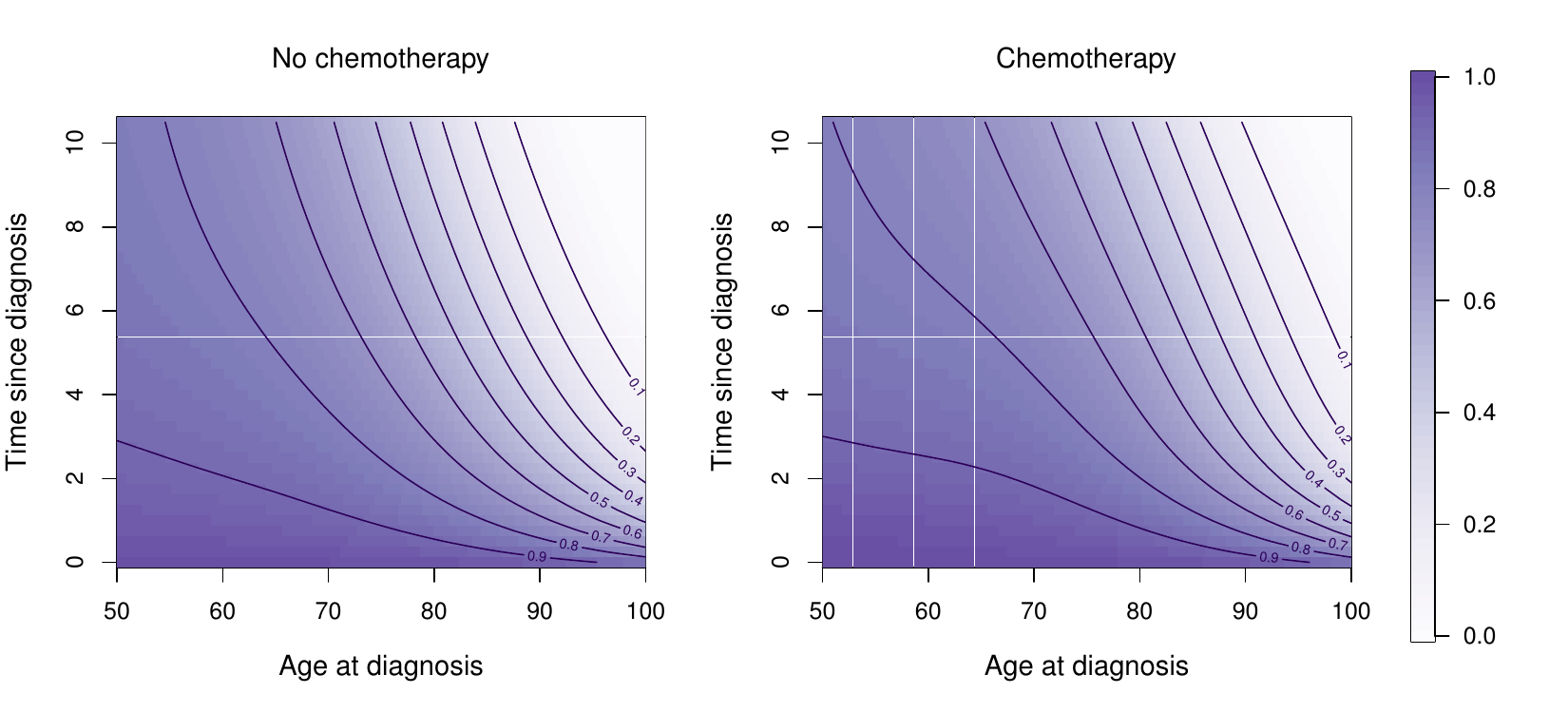} 
     \caption{Overall survival probabilities for the cause-specific hazards displayed in Fig.~\ref{fig:CSHwhiteOther}. }
    \label{fig:SurvwhiteOther}
\end{figure}

\end{document}